\documentclass[12pt]{article}
\usepackage{graphics}
\usepackage{float}
\usepackage{epsfig}
\usepackage{epstopdf}
\usepackage{setspace}
\title{Cosmology and thermodynamics of FRW universe with bulk viscous stiff fluid.}

\author{Titus K Mathew, Aswathy M B and Manoj M \\
Department of Physics, \\ Cochin University of Science and Technology, \\ Kochi-22,
India \\
E-mail: titus@cusat.ac.in}
\date{}
\begin{document}

\maketitle

\abstract{
We consider a cosmological model dominated by stiff fluid with a constant bulk viscosity. We classify all the possible cases of the universe predicted 
by the model and analyzing the scale factor, density as well as the curvature scalar. We find that when the dimensionless constant bulk viscous parameter is in the 
range $0 < \bar\zeta <6$ the model began with a Big Bang,  and make a transition form the decelerating expansion epoch to an accelerating epoch, 
then tends to the de Sitter phase as $ t\to \infty$. The transition into the 
accelerating epoch would be in the recent past, when $4<\bar\zeta<6.$ For $\bar\zeta>6$ the model doesn't have a Big Bang and suffered an increase in the fluid density 
and scalar curvature as the universe expands, which are eventually saturates as the scale factor $a \to \infty$ in the future. We have analyzed the model 
with statefinder diagnostics and find that the model is different from $\Lambda$CDM model but approaches $\Lambda$CDM point as $a \to \infty.$ We have also analyzed the 
status of the generalized second law of thermodynamics with apparent horizon as the boundary of the universe and found that the law is generally satisfied when 
$0 \leq \bar\zeta <6$ and for $\bar\zeta >6$ the law is satisfied when the scale factor is larger than a minimum value.} 



%
%
\section{Introduction}
\label{intro}
The recent discovery on the current acceleration of the universe form type Ia supernovae data \cite{Perlmutter1,Riess1,Hicken1} have shown that about 72$\%$ of the 
energy density of the universe is in the form of an exotic component, which is capable of producing negative pressure, called dark energy. Cosmological data from other 
wide range of sources, the cosmic microwave background radiation \cite{Komatsu1,Larson2}, baryon accoustic oscillations \cite{Percival1}, cluster gas fractions \cite{Samushia1} 
and gamma ray burst \cite{Wang21,Samushia2} have all confirming this conclusion. In the remaining part of the energy density, it was concluded even earlier to the 
discovery of dark energy that, 23$\%$ of it consists of weakly interacting matter called dark matter. The evidence for this is coming from variety of observational 
tests including weak \cite{Refregier1} and strong \cite{Tyson1} lensing, large scale structure \cite{Allen1}, as well from supernovae and cosmic microwave background 
\cite{Zwicky1,Rubin1,Rubin2}. In spite of the fact that all these observational data establishes the existence of the components like dark matter and dark energy, 
the existence of other exotic fluid components has not been ruled out. For example several models predicts the existence of an exotic component called dark radiation 
in the universe \cite{Dutta1}. Another exotic fluid which is predicted by several models is stiff fluid, a fluid with an equation of state $p_s=\rho_s,$ where $p_s$ 
and $\rho_s$ are the normal pressure and density of the stiff fluid respectively. The equation 
of state parameter of this fluid is seems to the largest value (equal to 1) consistent with causality, because the speed of sound in this fluid is equal to the speed 
of light.

The model with stiff fluid was first studied by Zeldovich \cite{Zeldovich1}. In recent years a large number of models have been proposed in studying the various 
cosmological properties of stiff fluid. In certain models with self-interacting dark matter components, the self interaction between the dark matter particles
is characterized by the exchange of vector mesons via minimal coupling. In such models the self interaction energy is shown to behave like a stiff fluid \cite{Stiele1}.
Stiff fluid is considered in certain cosmological models based on Horava-lifeshitz gravity. In Horava-lifeshitz gravity theories a ``detailed balancing'' condition was 
imposed as a convenient simplification and the usefulness of this detailed balancing condition was discussed in references \cite{Horava1,Calcagni1,Kiritsis1}. The 
stiff is fund to be arised in such models where this detailed balancing condition is relaxed \cite{Sotiriou1,Bogdanos1,Carloni1,Leon1}. Cosmological models with stiff 
fluid, based on Horava-Lifeshits gravity, have been studied in references \cite{Ali1,Dutta2}. The existence of stiff fluid have been found as exact non-singular 
solutions in certain inhomogeneous cosmological models \cite{Fernandez1,Dadhich1,Mars1,Fernanswz2}. The relevance of the stiff fluid equation of state to the matter 
content of the universe in the early stage of the universe was investigated in reference \cite{Barrow1}. The decrease in the density of stiff fluid in the universe is found 
to be faster than that of radiation and matter, hence it's effect on expansion would be larger in the initial stage of the universe. Primordial Nucleosysnthesis is 
an event took place in the early phase of the universe, a limit on the density of the stiff fluid can be obtained from big bang nucleosysnthesis constraint as in reference 
\cite{Dutta3}.

In a homogeneous and isotropic universe bulk viscosity is the unique viscous effect capable to modify the background dynamics. From a theoretical point of view, 
bulk viscosity arised in a system due to it's deviations from local thermodynamic equilibrium \cite{Zimdahl1}. In cosmology, bulk viscosity arised as an effective 
pressure, restoring the system to its thermal equilibrium, which was broken when universe expands too fast so that the system may not get enough time to restore the 
local thermal equilibrium \cite{Wilson1,Mathews1,Martens1,Okamura1}. Several years ago, before the discovery of the present acceleration of the universe, 
it has been proposed, in the context of the inflationary period of  early universe
that bulk viscous fluid can produce acceleration in the expansion of the universe
\cite{Heller1,Heller2,Waga1,Beesham1,Zimdahl3,Paddy1}. Recently investigations were made on possibility causing the recent acceleration of the universe with bulk 
viscous matter \cite{Avelino1,Avelino2}.

In the present work we study a stiff fluid dominated cosmological model with bulk viscosity. We assume a stiff fluid of equation of state $p_s=\rho_s$ 
and the bulk viscosity 
is characterized by a constant viscosity coefficient, which is the simplest parametrization for the bulk viscosity. We are deriving the Hubble parameter, density, 
equation of state, deceleration parameter and analyzing their behaviors for further possibilities including the recent acceleration of the universe. the paper is 
organized as follows. In section 2 we are giving the basic equations of FRW universe and deriving the general equation for the Hubble parameter in a bulk viscous stiff 
fluid dominated universe. We are classifying the different cases arised depending the value of the bulk viscous parameter and analyzing the evolution of various  
cosmological parameters. Section 3 containing the statefinder diagnosis of the model. In section 4, we presents the status of the generalized second law of 
thermodynamics followed by conclusions in section 5.

\section{Stiff fluid with bulk viscosity}

Stiff fluid cosmological models create interest because in these fluids the speed of light is equal to the speed of sound and its governing equations have the same 
characteristics as that of the gravitational field \cite{Wesson1}. The equation of state of the stiff fluid is given as \cite{Zeldovich1},
\begin{equation} \label{eqn:EOS1}
 p_s = \rho_s.
\end{equation}
This equation of state resembles the equation of state of a special case of models investigated by Masso and others \cite{Masso1}.

In cosmological models the effect of bulk viscosity can be shown to be an added correction to the net pressure $p_s^{'}$ as,
\begin{equation} \label{eqn:CP}
 p_s^{'} = p_s - 3 \zeta H,
\end{equation}
where $\zeta$ is the constant coefficient of viscosity and $H$ is the Hubble parameter. The form of the above equation was originally proposed by Eckart \cite{Eckart1} in the 
context of relativistic dissipative process occurring in thermodynamic systems went out of local thermal equilibrium. Later Landau and Lifeshitz proposed an 
equivalent formulation \cite{Landau1}. However Eckart theory has got the short comings that, it describes all the equilibria as unstable \cite{Hiscock1} and signals 
can propagate through the fluid with superluminal velocities \cite{Israel1}. Later Israel and Stewart proposed a more general theory which avoids these problems and 
from which Eckart theory is appearing as the first order limit \cite{Israel2,Israel3}. However because of the simple form of Eckart theory, 
it has been widely used by several authors to characterize the bulk viscous fluid. For example the Eckart approach has been used in models explaining the recent 
acceleration of the universe with bulk viscous fluid \cite{Kremer1,Cataldo1,Fabris1,Hu1,Ren1}. More over Hiscock et. al. \cite{Hiscock2} have shown that 
Eckart theory can be favored over the Israel-Stewart model, in explaining the inflationary acceleration of FRW universe with bulk viscous fluid. These motivate 
the use Eckart results, especially when one try to look at the phenomenon recent acceleration of the universe. At this point one should note the more general 
formulation than the Israel-Stewart by Pavon et. al. dealing with thermodynamic equilibrium \cite{Pavon1}

We consider the flat FRW universe favored by the recent WMAP observation \cite{WMAP1} with the scale factor
\begin{equation}
 ds^2 = -dt^2 + a^2(t) \left(dr^2 + r^2 d\theta^2 + r^2\sin\theta d\phi^2 \right),
\end{equation}
where $a(t)$ is the scale factor, $t$ is the cosmic time and $(r,\theta,\phi)$ are the comoving coordinates. The corresponding dynamics equations are,
\begin{equation} \label{eqn:frw1}
 H^2 = {\rho \over 3}
\end{equation}
\begin{equation} \label{eqn:frw2}
 2{\ddot{a} \over a} + \left({\dot{a} \over a}\right)^2 = p^{'}
\end{equation}
and the conservation equation,
\begin{equation} \label{eqn:con}
 \dot{\rho} + 3H \left(\rho + p^{'} \right) = 0,
\end{equation}
where we have adopted the standard units convention, $8\pi G=1$ and over-dot represent a derivative with respect to cosmic time.
 From the dynamical equations (\ref{eqn:frw1}) and (\ref{eqn:frw2}), we can formulate a first order differential 
equation for the Hubble parameter by using equations (\ref{eqn:EOS1}),(\ref{eqn:CP}) and (\ref{eqn:con}) as,
\begin{equation} \label{eqn:dotH}
 \dot{H}={3H\over 2} \left(\zeta -2H \right).
\end{equation}
The above equation can expressed in terms of the variable $x=\log a,$ suitably integrated and the final result can be written in terms of the scale factor as,
\begin{equation} \label{eqn:hp1}
 H = {H_0 \over 6} \left[\bar\zeta+\left(6-\bar\zeta\right) a^{-3} \right],
\end{equation}
where $\bar\zeta=3\zeta/H_0$ is the dimensionless bulk viscous coefficient, $H_0$ is the present value of the Hubble parameter and we have made the assumption 
that the present value of the density parameter of the stiff fluid $\Omega_{s0}=1$ for a stiff fluid dominated universe.

\subsection{Classification and evolution of the bulk viscous stiff fluid dominated model}

Equation for the Hubble parameter shows that for different value of the viscosity coefficient $\bar\zeta$ we get different models. In this section we are classifying 
different models of the universe arises due to the different values of the dimensionless viscosity coefficient. We analyse the behavior of the scale factor, density 
and other parameter of these different cases.

\subsubsection{Case-1: $\bar\zeta=0$}

 This corresponds to the universe dominated with stiff fluid without bulk viscosity. From equation (\ref{eqn:hp1}) the Hubble parameter becomes $H=H_0 a^{-3}.$ From 
the dynamical equation (\ref{eqn:frw1}) the corresponding density of the stiff fluid follows a relation,
\begin{equation} \label{eqn:rhobehav1}
 \rho_s \propto a^{-6}.
\end{equation}
This shows that the density of the non-viscous stiff fluid decays more rapidly than the non-relativistic matter or radiation in a FRW universe, which implies that 
the effect of the stiff fluid on the expansion of the universe would be the larger at early times. So the limit on the density of the stiff fluid can obtained by 
considering its effect the Big Bang nucleosysnthesis. Dutta et. al \cite{Dutta3} made an investigation in this regard and found that the change in the primordial abundance of 
helium-4 is proportional to the ratio $\rho_s/\rho_R,$ where $\rho_R$ is the radiation density. Consequently they found a limit on the non-viscous stiff fluid 
density as $\rho_s/\rho_R < 30$ when the temperature of the universe was around 10 MeV.

The evolution of the scale factor can be obtained by integrating the resulting Hubble parameter as,
\begin{equation}
 a(t) = \left(3 H_0 \left(t - t_0 \right) +1 \right)^{1/3}
\end{equation}
A second order derivative of the scale factor with time is,
\begin{equation}
 {d^2a \over dt^2} = -{2 H_0^2 \over \left(3H_o(t - t_0) +1 \right)^{5/3} }.
\end{equation}
This shows that the universe will undergo an eternal deceleration in this case.

The behavior of the density from equation (\ref{eqn:rhobehav1}) reveals that as the scale factor $a(t)\to 0,$ the density $\rho_s \to \infty.$ This implies the 
existence of singularity at the beginning of the universe. This can be further chequed by calculating the curvature scalar for a flat FRW universe using the equation 
\cite{Kolb1}
\begin{equation} \label{eqn:curvscalar}
 R = \left( {\ddot{a} \over a} + H^2 \right).
\end{equation}
Using the equation for the Hubble parameter and its time derivative it can easily shown that $R \sim H^2,$ which according to the equation $H=H_0 a^{-3}$ implies that the 
curvature scale $R \to \infty$ as $a \to 0$ at the origin, confirming the presence of the initial singularity. So it can be concluded that in this 
case the universe had a Big Bang. 
The time elapsed since the Big Bang, $t_B,$ is found to be 
\begin{equation}
 t_B = t_0 - {1 \over 3 H_0}.
\end{equation}
Also it is evident form the behavior of the density that, as $a(t) \to \infty$ the density $\rho_s \to 0.$ In this respect apart form the difference in the 
dependence on the scale factor, the general behavior of non-viscous stiff fluid is same as that of the non-relativistic matter or relativistic radiation.

\subsubsection{case-2: $0<\bar\zeta<6$}
\label{zeta}

\begin{figure}[h]
\centering
\includegraphics[scale=0.7]{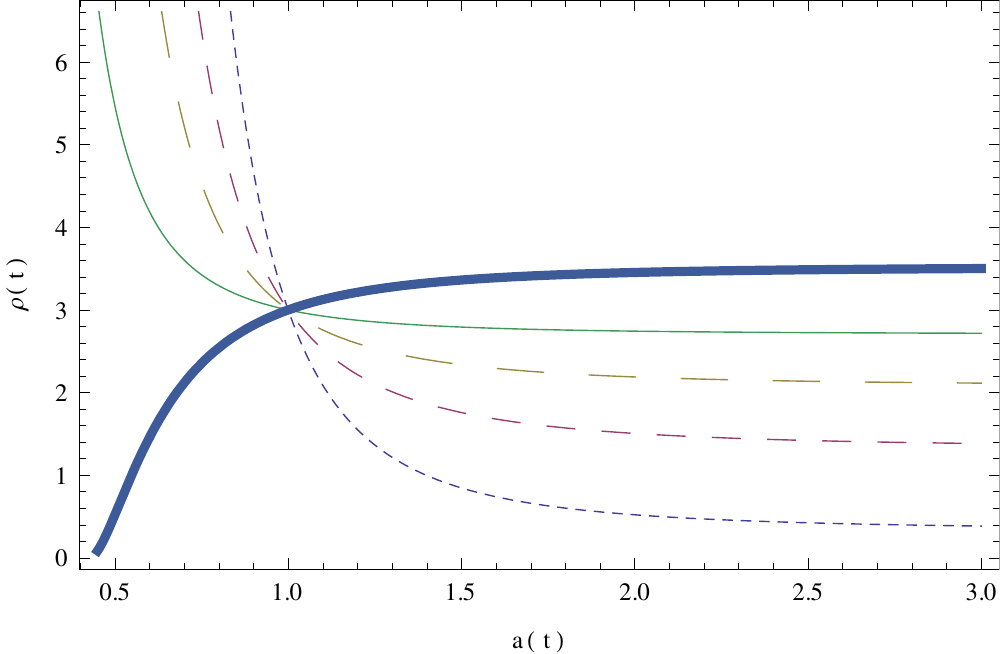}
\caption{Evolution of the density $\rho_s$ with scale factor $a(t).$ The bottommost line is for $\bar\zeta=2,$ the second from the bottom is for $\bar\zeta=4,$
the third line from the bottom is for $\bar\zeta=5,$ the fourth line is for $\bar\zeta=5.7$ and the topmost line is for $\bar\zeta=6.5.$}
\label{fig:density}
\end{figure}

The Hubble parameter is given by the equation (\ref{eqn:hp1}). Following the dynamical equations the density of the bulk viscous stiff fluid in this case is given as,
\begin{equation}
 \rho_s = 3 \left( \frac{H_0}{6} (\bar\zeta + (6 - \bar\zeta) a^{-3}) \right)^2
\end{equation}
The evolution of the scale factor is given in figure \ref{fig:density}

This equation shows that as the scale factor $a(t) \to 0$ the density $\rho_s \to \infty,$ indicating that there is a singularity at the origin. The presence 
of the singularity is further confirmed by calculating the curvature scale using equation (\ref{eqn:curvscalar}) and is,
\begin{equation} \label{eqn:curvscale2}
 R = {3H\bar\zeta \over 2} - H^2,
\end{equation}
 which shows that $R \to \infty$ as $a(t) \to 0,$ confirming the presence of the initial singularity. So the model of the universe in this case does have a Big Bang.

For finding the scale factor equation (\ref{eqn:hp1}) can be put in a form,
\begin{equation}
 {da^3 \over dt} - {H_0 \bar\zeta \over 2} a^3 = {H_0\left(6-\bar\zeta\right) \over 2},
\end{equation}
which can be suitably integrated for the scale factor as,
\begin{equation} \label{eqn:scalefact2}
 a(t)=\left({\bar\zeta - 6 + 6 \exp(\bar\zeta H_o [t-t_0]/2) \over \bar\zeta} \right)^{1/3}.
\end{equation}
This equation for scale factor reveals that, the time elapsed since the Big Bang is,
\begin{equation}
 t_B = t_0 + {2 \over H_0 \bar\zeta} \ln\left({6 - \bar\zeta \over 6} \right),
\end{equation}
hence the age of the universe since Big Bang is,
\begin{equation}
 t_0 - t_B = - {2 \over H_0 \bar\zeta} \ln\left({6 - \bar\zeta \over 6} \right).
\end{equation}
Taking $H_0 = 100 \, h \, km/sec/Mpc$, with $h=0.74$ the age of the universe is evaluated as per the above equation is around 13.8 Gyr for 
$\bar\eta=5.7,$ a value which is very closer to that predicted by the CMB anisotropy data \cite{Tegmark1}.

\begin{figure}[h]
\centering
\includegraphics[scale=0.7]{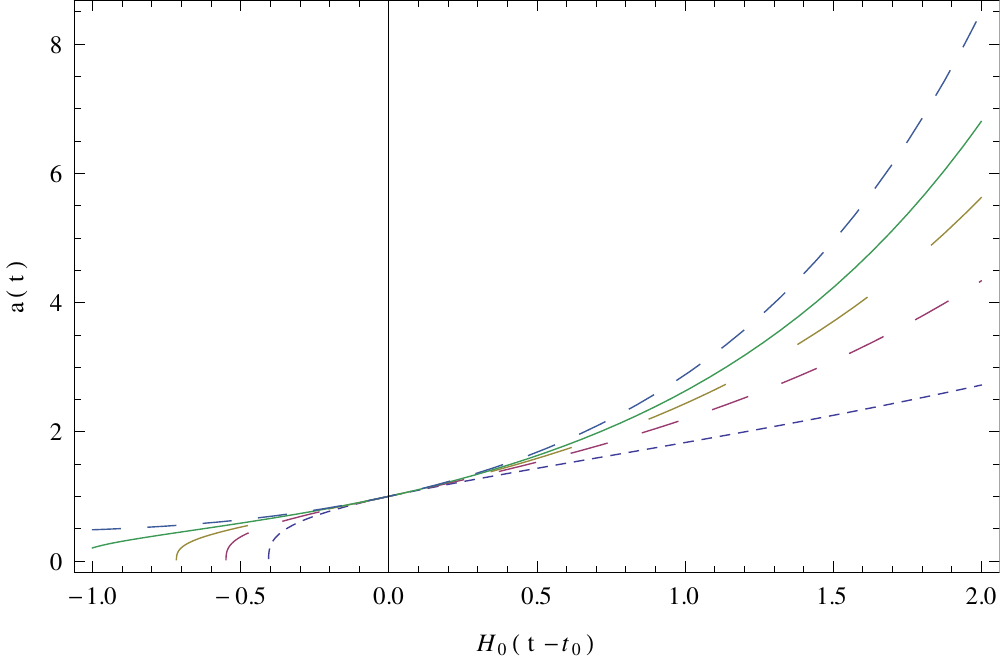}
\caption{ Evolution of the scale factor with $H_0(t-t_0).$ The lines represents, for $\bar\zeta=2,$ the bottommost line, for $\bar\zeta=4$
second line from the bottom, $\bar\zeta=5$ third form the bottom, $\bar\zeta=5.7$ fourth from bottom and $\bar\zeta=6.5$ the topmost line.}
\label{fig:a1}
\end{figure}

A plot of the evolution of the scale factor is given in figure \ref{fig:a1}. The scale factor equation (\ref{eqn:scalefact2}) 
shows that as $t \to \infty$ the scale factor approaches to a form like that of the de Sitter universe, 
\begin{equation} 
 a(t) \to \exp(\bar\zeta H_0[t-t_0]/2).
\end{equation}
While in the early stages of the evolution of the universe, when $\bar\zeta H_0 [t-t_0]/2 <1,$ the scale factor can be approximated as,
\begin{equation}
 a(t) \sim \left(1+3H_0 [t-t_0] \right)^{1/3}.
\end{equation}
These equations of the scale factor at the respective limits shows that the universe have an earlier deceleration phase followed by an acceleration 
phase in the later stage of the evolution. The value of the scale factor or redshift at which the transition form the decelerated to the accelerated expansion 
occur is depends on the viscosity coefficient $\bar\zeta$ as shown below. From the Hubble parameter we can calculate the derivative of $\dot{a}$ with respect $a(t)$ as,
\begin{equation}
 {d\dot{a} \over da} = \bar\zeta - 2 \left(6-\bar\zeta \right) a^{-3}.
\end{equation}
Equating this equation to zero, we obtained the transition scale factor as \cite{Avelino1},
\begin{equation} \label{eqn:trans-a}
 a_T = \left({2 (6 - \bar\zeta) \over \bar\zeta } \right)^{1/3},
\end{equation}
and the corresponding transition redshift is,
\begin{equation} \label{eqn:trans-z}
 z_T = \left({\bar\zeta \over 2(6 - \bar\zeta) } \right)^{1/3} - 1.
\end{equation}
From equations (\ref{eqn:trans-a}) and (\ref{eqn:trans-z}) it is clear that for $\bar\zeta=4$ the transition from the decelerated phase to the accelerated phase 
is occur at $z_T=0, \, a_T=1$ corresponds to the present stage of the universe. In the range $0<\bar\zeta<4$ the transition between the 
decelerated and the accelerated phase takes place in future corresponds to $z_T<0, \, a_T>1.$ The transition takes place in the past of the universe 
($z_T>0, \, a_T<1$) when $4<\bar\zeta<6.$ When $\bar\zeta = 0$ the value of $z_T$ becomes -1 and value of scale factor $a_T$ becomes infinity in the future, implies 
that no transition to accelerated expansion within a finite time and the universe is always decelerating.
While for $\bar\zeta =6$ the transition takes place at a time corresponds to $a_T \to 0$ closer to the Big Bang.

\begin{figure}
 \centering
\includegraphics[scale=0.7]{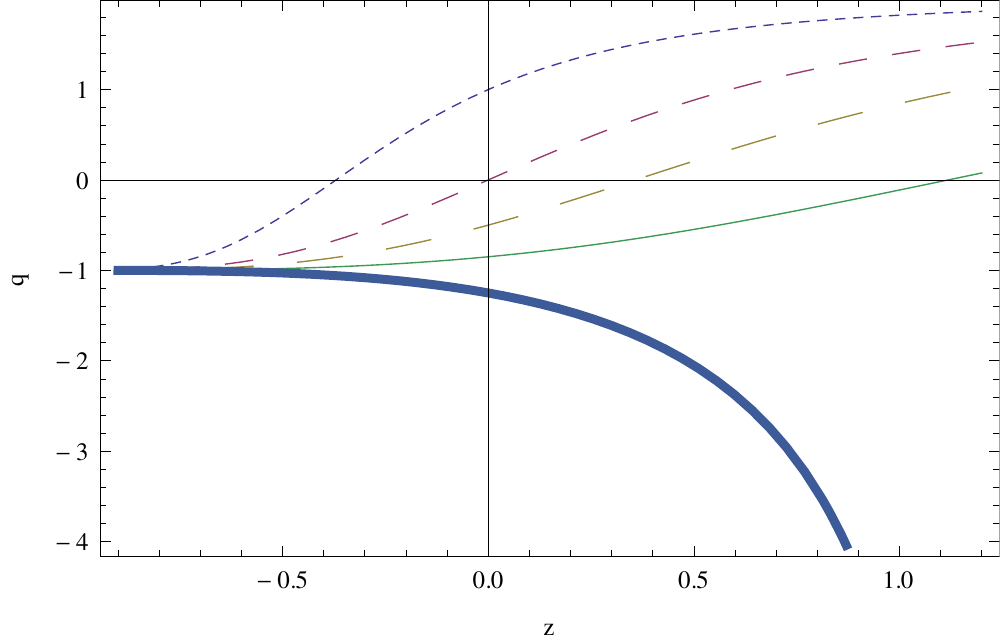}
\caption{ Evolution of the deceleration parameter with redshift The lines represents, for $\bar\zeta=6.5,$ the bottommost line, for $\bar\zeta=5.7$
second line from the bottom, $\bar\zeta=5$ third form the bottom, $\bar\zeta=4$ fourth from bottom and $\bar\zeta=2$ the topmost line.}
\label{fig:qpara}
\end{figure}

As a further clarification of the conclusions in the above paragraph we evaluate the deceleration parameter and the equation of state parameter of the bulk viscous 
stiff fluid in this case. A positive value of the deceleration parameter characterizes a decelerating universe, while a negative value characterizes an 
accelerating universe. The deceleration parameter $q$ can be evaluated using the equation,
\begin{equation} 
  q = -1 - {\dot{H} \over H^2}
\end{equation}
Using the Hubble parameter from equation (\ref{eqn:hp1}), the deceleration parameter in terms of the redshift $z$,
\begin{equation} \label{eqn:qparameter}
 q = -1 - {3 (\bar\zeta -6) (1+z)^3 \over \bar\zeta + (6-\bar\zeta) (1+z)^3 }
\end{equation}
where we took $a = (1+z)^{-1}.$
The evolution of the deceleration parameter is shown in figure \ref{fig:qpara}. It is clear from the figure that the deceleration parameter $q \to -1$ in the 
far future of the evolution of the universe as $z \to -1$ for any positive value of the dimensionless bulk viscous parameter $\bar\zeta.$

The transition redshift $z_t$ can be obtained by equating $q$ to zero, and it lead to equation (\ref{eqn:trans-z}). 
 For $\bar\zeta=0$ the deceleration parameter will be 2, corresponds to a universe dominated with non-viscous stiff fluid. For $\bar\zeta=6$ 
the parameter $q=-1$ corresponds to the de Sitter phase. So for $0<\bar\zeta<6$ the deceleration parameter is always decreasing form $q(a=0)=2$ to $q(a=\infty)=-1$, with 
a transition from positive to negative values corresponds to the transition from deceleration to acceleration in the expansion of the universe. The deceleration 
parameter for today, i.e. for $z=0$ is found to be,
\begin{equation}
 q(a=1) = 2 - \frac{\bar\zeta}{2}.
\end{equation}
This is agreeing with our earlier results in equations (\ref{eqn:trans-a}) and (\ref{eqn:trans-z}) that for $\bar\zeta=4$ the universe would enter the accelerating 
phase from the decelerated expansion at the present time. For $\bar\zeta<4,$ then $q>0$, we have decelerating universe today and for 
$\bar\zeta>4$ then $q<0,$ we have accelerating universe today. From the current observational results 
\cite{WMAP1,Tegmark1}, the present value of the deceleration parameter is around $-0.64\pm 0.03$, from which the bulk 
viscous coefficient is to be $\bar\zeta > 4$ 
for a universe dominated with bulk viscous stiff fluid. These analysis shows that a universe dominated with bulk viscous stiff fluid, it
 can take the role of the 
conventional dark energy, to cause the recent acceleration of the universe for a bulk viscous coefficient in the range $4<\bar\zeta<6.$

\begin{figure}
 \centering
\includegraphics[scale=0.7]{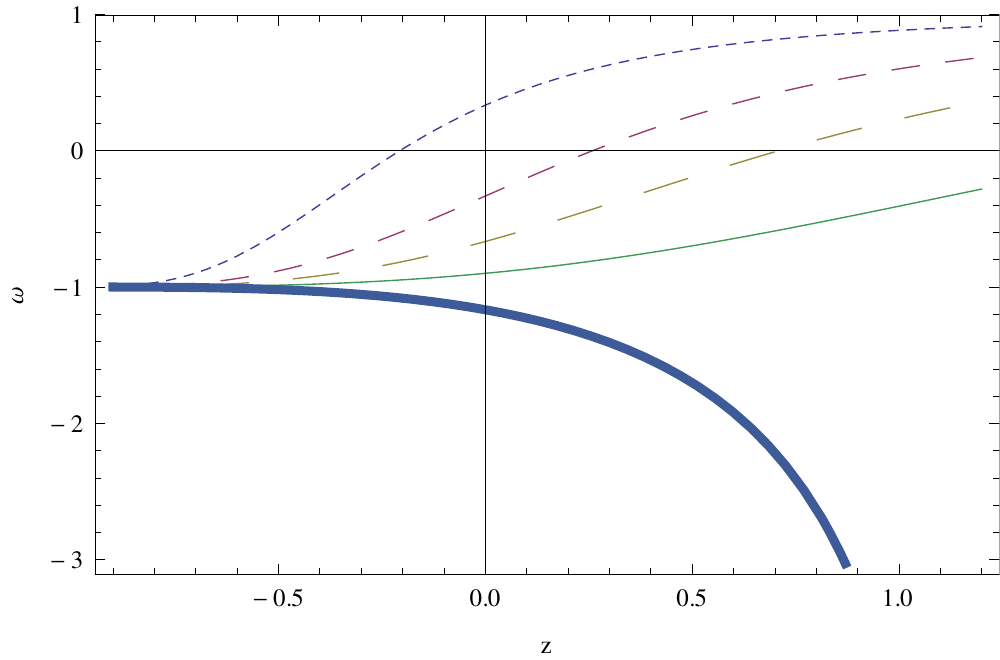}
\caption{ Evolution of the deceleration parameter with redshift The lines represents, for $\bar\zeta=6.5,$ the bottommost line, for $\bar\zeta=5.7$
second line from the bottom, $\bar\zeta=5$ third form the bottom, $\bar\zeta=4$ fourth from bottom and $\bar\zeta=2$ the topmost line.}
\label{fig:eos1}
\end{figure}

The evolution of the equation of state $\omega_s$ of the stiff fluid with bulk viscosity can be studied by calculating it using the relation \cite{tkm2},
\begin{equation}
 \omega_s = -1 - \frac{1}{3} {d \ln h^2 \over dx},
\end{equation}
where $h=H/H_0$ the weighted Hubble parameter. Evaluating $\omega_s$ in terms of the redshift $z$ gives,
\begin{equation}
 \omega_s = -1 - \left( { 2 (\bar\zeta - 6) (1+z)^3 \over \bar\zeta + (6 - \bar\zeta) (1+z)^3} \right).
\end{equation}
The evolution of the equation of state is as shown in figure \ref{fig:eos1}. 
As $z \to -1, \, (a \to \infty)$ the equation of state parameter $\omega_s \to -1$ in the future corresponds to the de Sitter universe, irrespective of the value of 
the viscosity coefficient. For $\bar\zeta=0$ the equation of state 
parameter become $\omega_s=1,$ implying the equation of state for the non-viscous stiff fluid, $p_s=\rho_s.$ For $\bar\zeta=6$ the $\omega_s$ become equal to -1. In 
the range $0<\bar\zeta <6$ the equation of state varies from $+1$ to $-1,$ and making a transition 
from positive to negative values. Event though negative value of $\omega_s$ leads to negative pressure, but for  universe to be in the accelerating 
phase, $\omega_s < -1/3.$
The present value of $\omega_s$ is found to 
be
\begin{equation} \label{eqn:omegapres}
 \omega_s (a=1)= 1 - \frac{\bar\zeta}{3}.
\end{equation}
This equation reveals that $\omega_s$ make a transition from positive values to negative values at the present time if $\bar\zeta = 3.$  While 
considering the evolution of the $q(a=1)$ parameter, we have shown that, $q$ make a transit to the negative values, giving a universe with accelerated expansion 
for $\bar\zeta=4.$ The negativity of $q$ parameter implies that the universe is accelerating and at the same time the equation of state parameter must be less 
than $-1/3$ for the universe to be an accelerated one \cite{Bamba1}. From equation (\ref{eqn:omegapres}) it is clear that $\omega_s(a=1) < -1/3$ 
only for $\bar\zeta \geq 4.$ The current observational value of equation of state parameter of the fluid responsible for the recent acceleration is around 
$-0.94\pm 0.1,$ \cite{Tegmark1} and form equation (\ref{eqn:omegapres}) we can 
infer that in a universe dominated with bulk viscous stiff fluid, the corresponding value of the bulk viscous coefficient 
is $\bar\zeta>4$ to  cause a recent acceleration.
So the analysis on the evolution of $\omega_s$ also shows that the bulk viscous stiff fluid can replace the conventional dark energy in causing the recent 
acceleration, for $4<\bar\zeta<6$.

\subsubsection{Case-3: $\bar\zeta>6$}

The equations (\ref{eqn:hp1}) and (\ref{eqn:scalefact2}) can be used in this case too for assessing the behaviors of the Hubble parameter and scale factor. 
For $\bar\zeta>6$ these equations shows that the resulting universe will always be accelerating. That is there is no decelerating epoch at all. When $t \to \infty$ the 
universe tends to the de Sitter phase.But when $t-t_0 \to -\infty$ the scale factor tend to finite minimum value (see figure \ref{fig:a1}) instead of zero  and is given as,
\begin{equation} \label{eqn:amin1}
 \lim_{ t-t_0 \to -\infty} a(t) \equiv a_{min}=\left(1 - \frac{6}{\bar\zeta} \right)^{1/3}
\end{equation}
The corresponding derivatives $\dot{a}$ and $\ddot{a}$ are zero, hence in this limit the universe become a Einstein static universe. As the universe evolves 
the scale factor increases monotonically. So there is no Big Bang in this case and the age of the universe is not properly defined.

The curvature scalar can be obtained using equation (\ref{eqn:curvscale2}). 
At  $a=a_{min}$, both $\ddot{a}$ and $H$ are zero, hence curvature scalar is also zero and it increases as the universe expands, attains the maximum value 
$R=\frac{5}{9}\left(H_0\bar\zeta \right)^2$ when $a \to \infty.$
The density of the bulk viscous stiff fluid follows same behavior as the curvature scale (see figure \ref{fig:density}), 
the density is zero when $a=a_{min}$ and attains the maximum value 
$\left(H_0 \bar\zeta \right)^2 / 12$ as $a \to \infty.$

\section{Statefinder analysis for $4<\bar\zeta<6$}

In the analysis in section \ref{zeta} we have concluded that there is a transition form decelerated expansion to 
accelerated one in the recent past when $4<\bar\zeta<6.$ in the past phase. 
This gives us hope in considering the discovery of the recent acceleration of the universe in the context of a universe dominated with bulk 
viscous stiff fluid. The behavior of the scale factor, $q$ parameter and equation of state all shows that 
\begin{figure}[h]
\centering
\includegraphics[scale=0.7]{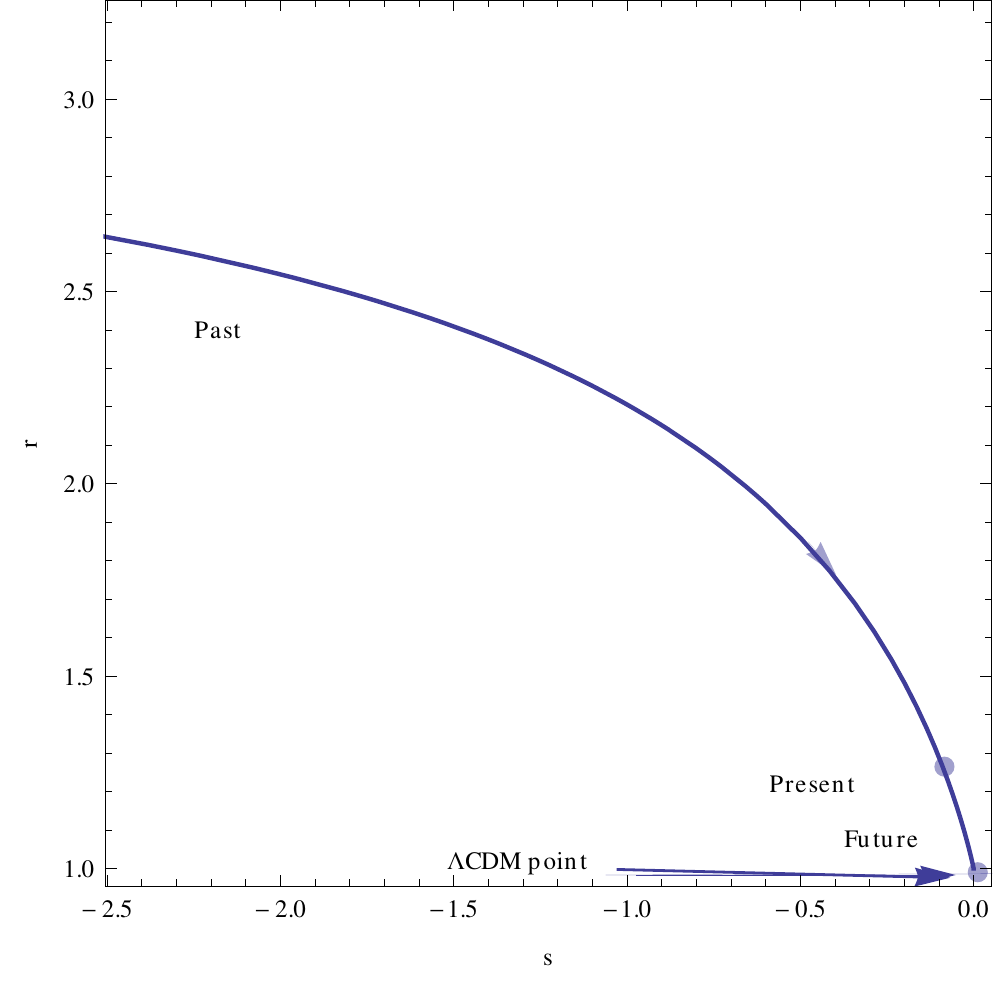}
\caption{The $r-s$ plane evolution of the model. The present position in the plane corresponds to values ($r_0,s_0)=(1.25,-0.08.)$ The evolution is the 
direction as shown by the arrow in the line.}
\label{fig:rsplot1}
\end{figure}
the bulk viscous stiff fluid is playing the role of dark energy. So we analyze the model using statefinder parameters to 
compare it with the standard dark energy models.
Statefinder parameters \cite{Sahni1} are sensitive tool to discriminate various dark energy models, and are defined as
\begin{equation} 
 r= {\ddot{H} \over H^3} - 3q -2
\end{equation}
and
\begin{equation}
 s={r - 1 \over 3(q - 1/2)}
\end{equation}
Using the equations for the Hubble parameter (\ref{eqn:hp1}) and deceleration parameter (\ref{eqn:qparameter}), the $r-s$ parameter equations can be expressed as,
\begin{equation} \label{eqn:r1}
 r={9 (6-\bar\zeta)^2 a^{-6} \over \left(\bar\zeta + (6-\bar\zeta) a^{-3} \right)^2 } + 1
\end{equation}
and 
\begin{equation} \label{eqn:s1}
 s= {2 (6 - \bar\zeta)^2 a^{-6} \over (6-\bar\zeta)^2 a^{-6} - \bar\zeta^2}
\end{equation}
The equations shows that in the limit $a \to \infty$ the statefinder parameters $(r,s) \to (1,0),$ a value similar to the $\Lambda$CDM model of the universe. 
A plot of the present model in the $r-s$ plane is shown in figure \ref{fig:rsplot1}, for bulk viscous coefficient $\bar\eta=5,$ and we found that the plot of other 
values of $\bar\zeta$ are also showing the same behavior, in fact the evolutions are coinciding each other. The plot reveals that the $(r,s)$ trajectory 
is lying in the region corresponds to $r>1 \, s<0,$ a character similar to that of generalized Chaplygin gas model of dark energy \cite{Wu1}. On the other hand in 
comparison with the holographic dark energy model with event horizon as the IR-cut-off \cite{Huang1,Wang4} whose $r-s$ evolution starts in the region $s\sim 2/3, \, r\sim 1,$ and end 
on the $\Lambda$CDM point, the present model starts in the region $r>1 \, s<0$ and end on the $\Lambda$CDM point in the $r-s$ plane.
Equations (\ref{eqn:r1}) and (\ref{eqn:s1}) shows that 
for $\bar\zeta=0,$ $(r,s)=(10,2)$ and for higher values of $\bar\zeta$ the $(r,s)$ parameter values decreases.  The values of the statefinder 
parameters for the present stage of the universe dominated with bulk viscous stiff fluid, 
corresponds to $a=1 \, (z=0)$ is,
\begin{equation}
 r=2 \left(1 - \frac{\bar\zeta}{6} \right)^2 \, \, \, s={(1-\bar\zeta/6) \over 3 (1-\bar\zeta/3)}
\end{equation}
This shows that as $\bar\zeta$ increases the present values of $(r,s)$ decreases. In figure \ref{fig:rsplot1} the present position of the universe is denoted and 
is corresponds to $(r,s)=(1.25,-0.08),$  which is different form the $\Lambda$CDM model, so the model is well discriminated form the $\Lambda$CDM model of the universe.

\section{Entropy and generalized second law of thermodynamics}

Bulk viscosity may be the only dissipative effect occurring in a homogeneous and isotropic universe. Any covariant description of dissipative fluids is subjected to the 
conservation equation,
\begin{equation} \label{eqn:tmunu1}
 T^{\mu\nu}_{;\mu}=0,
\end{equation}
provided there does not occur any matter creation, where the semicolon denote the covariant derivative and $T^{\mu\nu}$ is the energy momentum tensor of the fluid 
in the universe. The energy momentum tensor in covariant form is given as \cite{Weinberg1,Weinberg2},
\begin{equation}
 T_{\mu\nu}=\rho u_{\mu} u_{\nu}+(g_{\mu\nu}+u_{\mu} u_{\nu}) p^{'},
\end{equation}
where $u_{\mu}$ is the velocity of the observer who measures the pressure $p^{'},$ whose form is as given in equation (\ref{eqn:CP}). The conservation equation with 
the above form of the energy momentum tensor will lead to the equation (\ref{eqn:con}). The bulk viscosity causes the generation of local entropy in the FRW universe 
\cite{Weinberg1,Weinberg2}. The viscous entropy generation in the early universe was studied in reference \cite{Brevik2}. During the evolution of the universe the 
sum of the entropies of the fluid within the universe and that of the horizon must always greater than or equal zero, this is well known as the generalized second law 
(GSL) of thermodynamics. The satatus of the GSL for flat FRW universe with matter and cosmological vacuum was discussed in reference \cite{tkm1}. The status of the GSL 
in a flat universe with viscous dark energy was discussed in reference \cite{Karami2} and the authors have shown that the GSL is valid in FRW universe with apparent horizon as the 
boundary.

In this section we analyze about the validity of GSL in the present model of the universe dominated with bulk viscous stiff fluid by taking the apparent horizon 
as the boundary of the universe. The GSL can be formally stated as
\begin{equation} \label{eqn:gsl1}
 \dot{S}_s + \dot{S}_h \geq 0,
\end{equation}
where $S_s$ is the entropy of the stiff fluid and $S_h$ is that of the apparent horizon of the universe. The entropy of the stiff fluid within the horizon of the 
universe is related to its energy density and pressure through the Gibb's relation \cite{Izquierdo1},
\begin{equation}
 T dS_s = d(\rho_s V) + p^{'} dV,
\end{equation}
where $V=4\pi/3 H^3$ is the volume of the universe within the apparent horizon with radius $r=H^{-1}$ and $T$ is the temperature of the fluid within the horizon. 
We take the temperature $T=H/2 \pi$ equal to Hawking temperature of the horizon with the assumption that the fluid within the horizon is in equilibrium with the 
horizon, so there is no effective flow of the fluid towards the horizon. Using the dynamical equation and the net pressure in equation (\ref{eqn:CP}),
the time evolution of the entropy of the bulk viscous stiff fluid within horizon become,
\begin{equation} \label{eqn:Ss}
 \dot{S}_s = {16\pi^2 \dot{H} \over H^3} - {24\pi^2 \dot{H} \over H^4} \left(2 H - \bar\zeta \right)
\end{equation}

The entropy of the apparent horizon is given by the Bakenstein-Hawking formula \cite{Beken1,Hawking1,Davies3},
\begin{equation}
 S_h = 2\pi A
\end{equation}
where $A=4\pi H^2$ is the area of the apparent horizon. Hence the time rate of the horizon entropy is,
\begin{equation} \label{eqn:Sh}
 \dot{S}_h = -{16 \pi^2 \dot{H} \over H^3}.
\end{equation}
From the equations (\ref{eqn:Ss}) and (\ref{eqn:Sh}) the GSL condition equation (\ref{eqn:gsl1}) is satisfied if 
\begin{equation}
 \dot{H} \left(\bar\zeta - 2 H \right) \geq 0.
\end{equation}
Using the equation (\ref{eqn:dotH}) the above condition become,
\begin{equation}
 H \left(\bar\zeta - 2 H \right)^2 \geq 0.
\end{equation}
\begin{figure}[h]
 \centering
\includegraphics[scale=0.7]{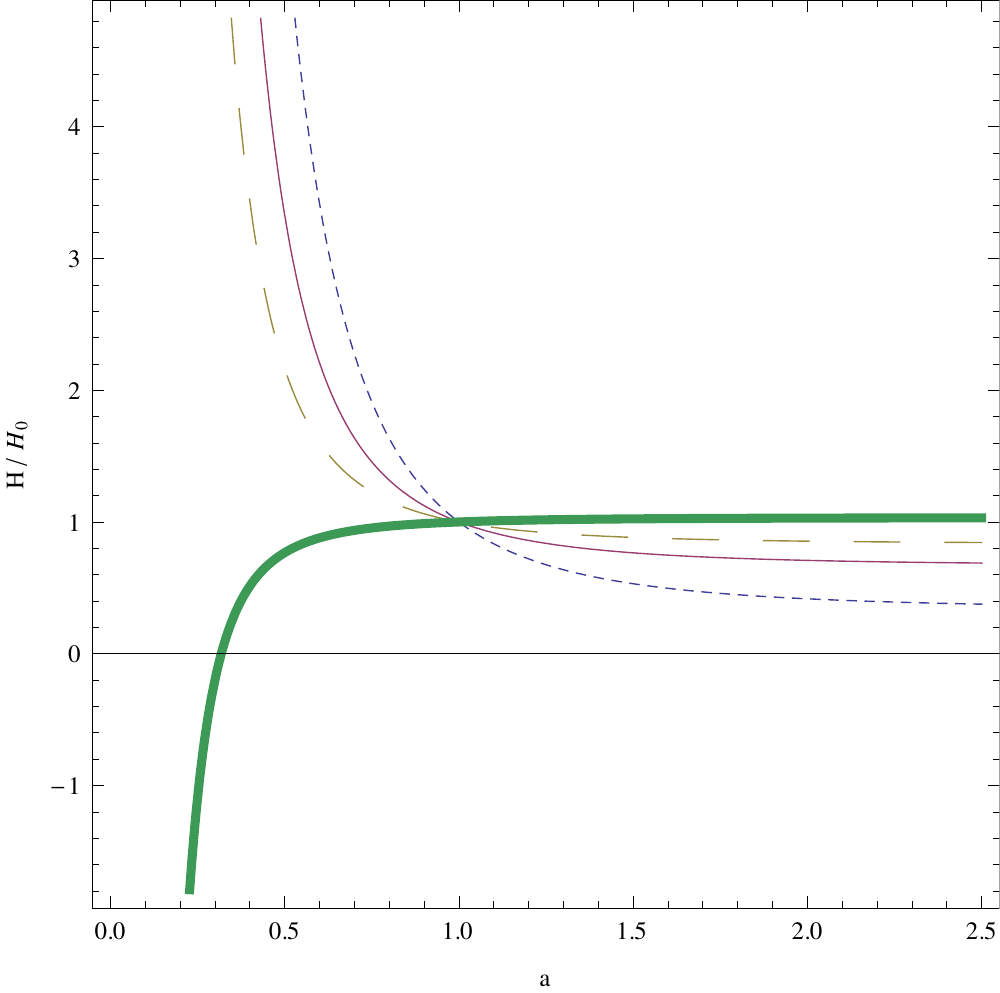}
\caption{ Evolution of the Hubble parameter with scale factor. Bottom line is for $\bar\zeta=2,$ second from bottom for $\bar\zeta=4,$
third from bottom is for $\bar\zeta=5$ and top thick line corresponds to $\bar\zeta=6.2.$}
\label{fig:hubble1}
\end{figure}

As far as $H$ is positive in an expanding the universe, it is evident that the GSL is satisfied in a bulk viscous stiff fluid dominated universe with 
apparent horizon as boundary. From equation (\ref{eqn:hp1}), the required condition for the validity of GSL is,
\begin{equation}
 \left(\bar\zeta + (6 - \bar\zeta) a^3 \right) \geq 0.
\end{equation}
For $\bar\zeta \leq 6$ the above condition is fullfiled consequently the GSL is well satisfied. But when $\bar\zeta>6$, the above condition is satisfied only when 
$a \geq a_{\min}$ given by equation (\ref{eqn:amin1}) and it is clear form the plot, fig. \ref{fig:hubble1} of the Hubble parameter with the scale factor.

\section{Conclusions}

In this paper we present a study of the bulk viscous stiff fluid dominated universe model with a constant bulk viscous coefficient $\bar\zeta.$ Stiff fluid is 
an exotic fluid with equation of state parameter $\omega_s=1,$ first studied by Zeldovich \cite{Zeldovich1}. We analyzed the different possible phases of the model 
according the value of the dimensionless bulk viscous parameter $\bar\zeta$ and we take $\bar\zeta \geq 0.$ 
For $\bar\zeta \geq 0$ the model predicts expanding universe in general. For $\bar\zeta=0$ the 
model reduces to non-viscous stiff fluid dominated universe began with a Big Bang, and is always decelerating with the density varying as $\rho \sim a^{-6}$ .

For $0<\bar\zeta<6$ the model is corresponds to a universe started with a Big Bang and undergoing a decelerated expansion first followed by a transition to the 
accelerated phase of expansion at later time. For $\bar\zeta=4$ the transition form the decelerated to accelerated expansion epoch takes place today. 
For $0<\bar\zeta<4$ the transition to the accelerated expansion phase is takes place in the future, but for $4 <\bar\zeta <6$ this transition is found to 
occur in the past. This shows that the bulk viscous stiff fluid can cause the recent acceleration of the universe. 
 From the behavior of the scale factor we have obtained the age of the universe as 
$t_0-t_B=-(2/H_0 \bar\zeta) \ln (1 - \bar\zeta/6).$ 

We have also studied the evolution of the deceleration parameter $q$ and the equation of state parameter $\omega_s$ for $0<\bar\zeta <6.$ For $\bar\zeta=4$ the 
deceleration parameter enter the negative region today, corresponding to accelerated universe at present. For $\bar\zeta < 4$ the $q$ enter the negative region in the 
future, while for $4<\bar\zeta <6$ it would enter the negative region in past, implying that the universe make a transition form the decelerated to its 
accelerated phase in the past. In general the 
$q \to -1$ as $a \to \infty,$ corresponds to de Sitter model of the universe. However for $\bar\zeta>6,$ $q$ ia always negative, implying eternal acceleration 
without any transition from the decelerated to accelerated epoch. 

Behavior of $\omega_s$ shows that, it's value is changing from positive to negative when $0<\bar\zeta<6$ implies a transition from decelerated to accelerated 
epoch and always negative when $\bar\zeta>6$ implies an eternal accelerated universe. But irrespective of the value of the viscous coefficient $\omega_s \to -1$ as 
$z \to -1 \, (a \to \infty).$ The equation of the today's value of $\omega_s$ indicating that it would be negative at present if $\bar\zeta>3,$ however that 
does not corresponds to an accelerating universe. For an accelerating universe $\omega_s <-\frac{1}{3}$ for which $\bar\zeta>4$ according to the equation 
of today's value of $\omega_s.$ 

Statefinder analysis of the model for $4<\bar\zeta<6$ were done, in which range the model predicts recent acceleration of the universe. The today's position of the 
model in the $r-s$ plane is found to be $(r_0,s_0)=(1.25,-0.08)$ and different from the $\Lambda$CDM model. However as $a \to \infty$ the statefinder parameters 
$(r,s) \to (1,0)$ corresponds to the $\Lambda$CDM point.

When $\bar\zeta>6$ we have found that as $(t_0 -t) \to -\infty$ the scale factor tends to a minimum, i.e. $a \to a_{min},$ given by equation (\ref{eqn:amin1}), and 
in this case the model doesn't have a Big Bang. The density and the curvature scalar are increasing as the universe expands and attains maximum as $a \to \infty.$

We have analyses the status the GSL in the present model, and found that the GSL of thermodynamics is generally valid with apparent horizon as the boundary when 
$0<\bar\zeta<6.$ But when $\bar\zeta>6$ The GSL is satisfied only if the scale factor, $a > a_{min}$, where $a_{min}$ is given by equation (\ref{eqn:amin1}).

Summerising the results, for $\bar\zeta=0$ the model reduces the stiff fluid dominated universe without viscosity. For $0<\bar\zeta<6$ the model predicts a universe 
with a Big Bang and make transition form the decelerated to the accelerated phase during it's evolution. For $\bar\zeta>6$ the model doesn't have a Big Bang, hence 
age is not properly define and the density and the curvature scalar increases as the universe expands and attains a maximum as $a \to \infty.$


\end{document}